# *in situ* surface-enhanced Raman spectroscopy to investigate polyyne formation during pulsed laser ablation in liquid

P. Marabotti, S. Peggiani, A. Facibeni, P. Serafini, A. Milani, V. Russo, A. Li Bassi, C. S. Casari[*]

Department of Energy, Micro and Nanostructured Materials Laboratory - NanoLab, Energy, Politecnico di Milano, Via Ponzio 34/3, Milano 20133, Italy

**Abstract**

The synthesis of polyynes during their formation by pulsed laser ablation in liquid (i.e., acetonitrile) has been analyzed by *in situ* surface-enhanced Raman spectroscopy (SERS). A polyethylene pellet, functionalized with silver nanoparticles and placed into the ablation medium, served as SERS active surface. This innovative approach granted the possibility to investigate the dynamics of formation and degradation of polyynes with a time-resolution of a few seconds, starting from the early stages of ablation when the concentration is low. The processes occurring during the synthesis have been studied comparing the *in situ* SERS signal of polyynes and byproducts in the solution. The different kinetics of short and long polyynes have been investigated by their *in situ* SERS signal, exploring the final distribution of chain lengths. *Ex situ* UV-Vis and high-performance liquid chromatography confirmed the observations gained from *in situ* SERS data and validated this innovative *in situ* and *in operando* analysis.

## 1. Introduction

Polyynes are one-dimensional systems composed by the alternation of single and triple bonds of sp hybridized carbon atoms [1,2], approaching, as the length increases, the ideal novel allotrope carbyne [3]. During the last decades, their appealing predicted optical, mechanical, and electronic properties, depending on the length and the chemical termination, have attracted the interest of material scientists and chemists [1,2,4]. Polyynes have been observed in interstellar media and natural compounds [5–8], and they can be synthesized in the laboratory by both chemical and physical methods [9–21], including the recently reported cyclo[18]carbon [22]. Among the physical methods, pulsed laser ablation in liquid (PLAL) has been widely employed in this field, thanks to the high efficiency, versatility, and simplicity of the technique [1,14,15,23–26]. PLAL consists of the irradiation of a carbon target immersed in a solvent by short laser pulses, usually in the ns range. The proper choice of the solvent and the laser wavelength granted the synthesis of the longest polyyne ever produced by physical methods (up to 30 carbon atoms) and with a variety of terminating groups [15,26–28]. Nevertheless, the mechanisms behind the synthesis of polyynes by PLAL have not been fully understood. The growth of carbon chains is characterized by the competition between two phenomena: polymerization reactions, which tend to extend the chain length, *versus* hydrogenation reactions, that instead terminate the chain (usually with hydrogen atoms) [23]. The synthesis has a radical nature since polymerization reactions are thought to happen by the addition of carbon dimers $C_2$ and/or ethynyl radicals [23,25]. The physics of PLAL suggests that the formation of polyynes occurs at the interface between the ablation plasma plume and the solvent, where strong out-of-equilibrium conditions take place and radicals are formed. However, there are no direct proofs of the processes taking place during the ablation since the short time scales involved (from *ps* to tens of *ns* range [29,30]) and the complex environment make it difficult to employ suitable diagnostic techniques. Indeed, *in situ* experiments could help to deepen the knowledge of what is happening during the formation of polyynes in solution.

---

[*] Corresponding author. Tel: +39 02 2399 6331. E-mail: carlo.casari@polimi.it

The optical and vibrational properties of polyynes have been studied by different techniques [1,31]. UV-Vis absorption spectroscopy is widely employed since polyynes possess characteristic vibronic patterns in the UV region that depends on both length and termination, allowing immediate detection of sp-carbon chains and the calculation of their concentration in solution [14,15,32]. Absorption spectroscopy can be used as a detection technique in high-performance liquid chromatography (HPLC) which allows a more sensitive detection, quantification, and selective separation of polyynes by their structure [10,15,26,27]. Moreover, the vibrational properties of polyynes have been extensively investigated by Raman and IR spectroscopy [1,31,33–36]. In particular, sp-carbon chains feature a distinctive Raman active collective CC stretching mode, called ECC mode from the "effective conjugation coordinate" model, located in the 1800-2200 $cm^{-1}$ spectral region where there are no other features from all the other carbon-based solids and nanostructures [31,37,38].

The elusive nature of carbyne is related to the poor stability usually suffered by sp-carbon chains since they tend to reorganize into more stable carbon allotropes, i.e. $sp^2$ or amorphous carbon species, *via* crosslinking reactions [1,15,21,31,39]. Termination with bulky endgroups and encapsulation in carbon nanotubes or in polymeric matrices allowed to improve their stability [1,3,16,40–42]. Another stabilization strategy is the addition of metal nanoparticles [40,42–44]. Moreover, the interaction with metal nanoparticles (usually gold or silver) boosts the Raman response of polyynes up to six orders of magnitude, the so-called surface-enhanced Raman scattering (SERS), if the Raman laser matches the surface plasmon frequency of the metal nanoparticles [31,45]. In this way, it is possible to study very low concentrated samples, that otherwise cannot be probed due to the weak or negligible Raman signal when the amount of probed material is scarce. SERS has been reported as a reliable method to detect polyynes in solution and in solid-state, and it appears a viable route for *in situ* monitoring of the formation process, thanks to its enhanced sensitivity [43,45,46]. However, the use of SERS to investigate vibrational properties of polyyne implies two main drawbacks: first, the sharp Raman ECC peak broadens into a large band given by the contribution of every possible geometric configuration of the complex polyyne-nanoparticle and by the usual wide size-distribution of chemically synthesized colloidal nanoparticles; second, the interaction between sp-carbon chains and metal nanoparticles introduces new vibrational modes that generate an intense additional feature, indicated as low-frequency (1800-2000 $cm^{-1}$) band, which may overlap with the characteristic vibrational peaks of polyynes [33,43,45,47,48]. In this respect, the application of SERS to investigate polyyne formation during the ablation process raises further non-trivial issues. First, metal nanoparticles suspended in the solution may absorb the laser pulse preventing its arrival at the target, thus reducing the ablation efficiency, and may subsequently be fragmented in smaller nanoparticles with different surface plasmon frequency, possibly reducing the SERS effect [30,49]. Second, metal nanoparticles in the solution may modify the formation process of polyynes, catalyzing chemical reactions, undergoing aggregation, and interacting with the growing sp-carbon chains, leading to pseudocarbynes [42,45,46,50,51].

Here, to overcome the above-mentioned limitations, we employed an innovative *in situ* SERS approach to analyze the processes occurring during the synthesis of polyynes by PLAL in acetonitrile. We selected acetonitrile because it proved to be an efficient liquid environment for the synthesis of polyynes and their stability [15,28,52]. Indeed, it was demonstrated that a high concentration of polyynes can be obtained in low polar solvents, as acetonitrile [15,53]. Moreover, polyynes are more stable in a liquid environment characterized by low polarity and a low quantity of dissolved oxygen [15]. Finally, acetonitrile is compatible with *ex situ* HPLC analyses that will be compared to the *in situ* SERS data. The novel SERS apparatus is composed of a silver nanoparticle-functionalized polyethylene (PE) pellet attached to the sides of the glass vial used for the ablation. In this way, no metal nanoparticles are suspended in the solution while ensuring high effective surface for SERS analysis. The SERS enhancement enables the detection of very low concentrations of polyynes, down to $10^{-8}$ M, allowing to follow the synthesis from the early formation stages. In summary, our approach provides a time-resolved investigation of the SERS signal of both polyynes and $sp^2$ carbon during the ablation process, granting the examination of the efficiency of the ablation and the possible degradation mechanisms of polyynes in solution. The broad observed polyyne band is deconvoluted with an *ad hoc* algorithm to track the different progression of short and long polyynes. The interpretation of *in situ* SERS data is supported by *ex situ* HPLC and UV-Vis measurements of liquid samples after specific ablation times, as well as DFT calculations of Raman activities of polyynes.

## 2. Material and methods

Pulsed laser ablation was carried out using the second harmonics ($\lambda$ = 532 nm) of an Nd:YAG pulsed laser (Quantel Q-Smart 850), with a pulse duration of 6 ns and a repetition rate of 10 Hz. The laser beam was focused using a plano-convex lens (200 mm focal length, Thorlabs). A graphite disc (8 mm diameter, 2 mm thick, Testbourne Ltd, purity 99.99%) was employed as an ablation target in a glass vial filled with 2 mL of acetonitrile (here denoted by ACN, Sigma-Aldrich, purity ≥ 99.9%). The fluence on the graphite target was set to 0.37 J/cm$^2$, calculated using the method reported in Eq. 3 in the SI. This method allowed us to account for the presence of the liquid meniscus in the calculation of the spot radius of the laser beam.

A polyethylene (PE) pellet functionalized with silver nanoparticles (AgNPs) was employed as a SERS substrate. The PE pellets were fabricated following the method described elsewhere [54]. The pellet is anchored to the sides of the vial through a polyvinyl alcohol layer as reported in Fig. 1a, dried in an oven at 90 °C. The target-pellet distance was chosen in the range 10 ± 2 mm. SERS spectra were continuously collected during all the ablation process employing a Renishaw inVia Raman microscope with a diode-pumped solid-state laser (λ = 532 nm). The laser power was set to 0.7 mW to avoid degradation of the pellet. The remote Raman probe is equipped with a 20x objective that focuses the laser beam down to a spot size of 12 μm. Each measure consists of 10 acquisitions of 1 s of exposure time. The selected grating of 1800 l/mm allows us to achieve a resolution of 9 cm$^{-1}$ and a spectral range of approximately 1650 cm$^{-1}$ in static acquisition mode. The remote Raman head is mounted on a tripod equipped with a micrometric translation stage. The Raman laser beam was focused onto the pellet from the outside of the glass vial. We employed the same laser wavelength for both PLAL and SERS and we verified that the edge filter of the Raman spectrometer can prevent the diffused pulsed laser beam of PLAL from entering the spectrometer and blinding the CCD. The surface of the PE pellet before and after the ablation was analyzed by scanning electron microscope (SEM, Zeiss Supra 40) equipped with a detector of secondary electrons, using an accelerating voltage of 3 kV.

The concentration of polyynes and the purity of the mixtures were analyzed by *ex situ* UV-Vis absorption spectroscopy (Shimadzu UV-1800 UV/visible Scanning Spectrophotometer, 190-1100 nm spectral range). All the solutions were diluted with pure ACN to avoid saturation of the most concentrated chains, with a proportion of 1/37 v/v. Quartz cuvettes with an optical path of 1 cm were employed and pure ACN was used as a reference in all the measurements.

The exact concentration of hydrogen-capped polyynes contained in the solutions was evaluated using reversed-phase high-performance liquid chromatography (RP-HPLC). A C18 column was used (Phenomenex Luna 3 μm C18(2) 100 Å, LC Column 150 x 4.6 mm), mounted on a Shimadzu Prominence UFLC, equipped with a photodiode array (DAD) UV-Vis spectrometer and a fraction collector module. SERS measurements of size-selected H-capped polyynes separated by HPLC were done in liquid employing the same PE pellets as probes. The separation was performed employing a gradient mode, whose mobile phase gradually varies in time by changing the amount of ACN (%B), and an overall flux of 0.8 mL/min. Before the HPLC analysis, the solutions were filtered through Phenomenex Phenex RC-membrane syringe filters (450 nm of pore size) to remove any possible macroscopic impurity that could damage the HPLC column.

Theoretical calculations based on state-of-the-art Density Functional Theory (DFT) simulations were performed at the PBE0/cc-pVTZ level of theory on single linear chains using Gaussian09[55] to support experimental results. Frequency analysis has been performed for H-capped polyynes with different lengths (HC$_{2n}$H, n=3-10). All the optimized molecules resulted to be linear. Then, prediction of the Raman spectra has been carried out for the optimized molecules and, in particular, Raman activities were collected. Theoretical simulations made at this level of theory demonstrated from previous works to provide a very good agreement with experiments[1,31].

## 3. Results and discussion
### 3.1. A novel approach for *in situ* SERS during laser ablation in acetonitrile
To continuously perform *in situ* SERS measurements during the ablation, we employed the experimental setup reported in Figure 1a. SERS effect, thanks to the enhancement of the Raman signal, allows us to examine the formation of polyynes with time-resolution of a few seconds from the early stages of ablation, when the concentration of polyynes is very low, or in the case of long chains that are always too low concentrated for conventional Raman. We employed a solid-state fixed SERS active surface to avoid the use of colloidal nanoparticles in liquid which may modify the liquid environment and the formation process of polyynes. Thus, the core of our *in situ* SERS approach is a PE pellet functionalized with silver nanoparticles, realized following a patented method [54]. The morphology of the SERS active pellet was investigated by SEM before and after the ablation (Fig. 1b and 1c) to inspect AgNPs distribution and mean size. We estimated the average nanoparticle size of about 90 nm, from their plasmonic resonance, peaked at about 475 nm [56]. The interaction of polyynes with silver nanoparticles generates an additional surface plasmon absorption at a higher wavelength, as observed in other works [45,57]. Indeed, the pellet shows optimal SERS response by exploiting a Raman laser beam at 532 nm, i.e. the same laser wavelength used for PLAL synthesis. SEM images (see Fig1.b) show that the distribution of AgNPs is uniform compared to the spot size of the Raman laser (12 µm). Furthermore, the dimension of the Raman laser is sufficient to cover a large number of silver nanoparticles, ensuring a SERS response that does not depend on the specific point where the Raman laser beam is focused.

To explore the phenomena taking place during polyynes synthesis by PLAL, we employ this novel SERS probe. The selected target-pellet distance (10 ± 2 mm) ensures no interactions between the plasma plume generated by the ablation and the PE pellet. Indeed, it was reported in the literature that the maximum dimension of the plume and the subsequent bulkier cavitation bubble in PLAL experiments are limited below ca. 3 mm [58–62]. In these cases, the fluence (tens of $J/cm^2$) is much larger compared to that employed in this work (0.37 $J/cm^2$). So, we expect that the plasma plume and the cavitation bubble generated in our setup are further confined and do not interact with the PE pellet. After the laser pulse, we can distinguish between a fast transient and a slow relaxation phase. The fast transient corresponds to the shockwave lifetime (≤ 1 ms [29,30]), generated by the collapse of the cavitation bubble, during which the carbon species produced, i.e. polyynes and byproducts (mainly consisting of $sp^2$ carbon), spread out in the whole liquid volume and interact with the PE pellet. We assumed the initial speed of the carbon species equal to that of the shockwaves, i.e. approximately 1500 m/s [58,62,63]. For this reason, the *in situ* SERS measurements are unaffected by the position of the PE pellet if it is far enough from the plasma plume and within the volume reached by the shockwave before the subsequent laser pulse (after 100 ms in our setup, repetition rate of 10 Hz). After the shockwave, carbon species move in the solution in a Brownian-like motion [64,65] and can interact with the PE pellet and with each other, until the next laser pulse after 100 ms.

Once polyynes have reached the PE pellet, we assume that they stick to the AgNPs adsorbed on its surface and remain attached. To support this mechanism, we tested the adhesion of AgNPs to the pellet and the interaction between polyynes and AgNPs. Regarding the stability of AgNPs on the pellet, from SEM analysis after the ablation (see Fig. 1c), we noticed that the distribution of AgNPs remains unchanged, as well as the average size of the nanoparticles. Moreover, no silver nanoparticles were detected in the solution by UV-Vis and SERS analysis after 60 minutes of ablation, as confirmed by absorption spectra (see Fig. S1 in the SI), and by the absence of SERS signal of carbon compounds by focusing the Raman laser inside the solution (see Fig. S2a). Thus, we assume that AgNPs are strongly attached to the PE pellet and do not suffer any detectable degradation or detachment process. Regarding the interaction of polyynes with AgNPs, we analyzed a PE pellet extracted from the solution after the ablation and left to dry in the air. The corresponding SERS spectrum clearly shows the presence of polyynes bonded to the AgNPs, together with a non-negligible fraction of $sp^2$ carbon (see Fig. S2b in the SI). We further tested the interaction between polyynes and AgNPs by sinking a PE pellet, covered by polyynes and $sp^2$-carbon byproducts from previous ablations, in pure acetonitrile solution, i.e. without carbon species dissolved in it. We did not observe any trace of polyynes in the acetonitrile solution (see Fig. S3 in the SI) in *ex situ* HPLC analysis, whose lower detection limit is approximately $10^{-9}$ mol/L for $C_8$. Thus, this suggests that polyynes are strongly bound to AgNPs on the PE pellet. In this way, we confirmed the stabilization of polyynes that usually degrade if dried at ambient condition and their strong interaction with the AgNPs. Indeed, it was already demonstrated that polyynes interact with AgNPs. This interaction is supposed to happen at the edges of the chain, characterized by a strong chemical interaction [15,33,45]. Together with providing SERS enhancement (up to $10^6$), this also contributes to stabilizing the chains, hindering their degradation – at least with colloidal nanoparticles [20,39,43,52]. Furthermore, by increasing the number of pellets attached to the vial walls and keeping fixed the ablation parameters, the overall concentration of polyynes in the solution decreases as shown in Fig. S1 in SI, meaning that polyynes dispersed in the solvent are captured by AgNPs.

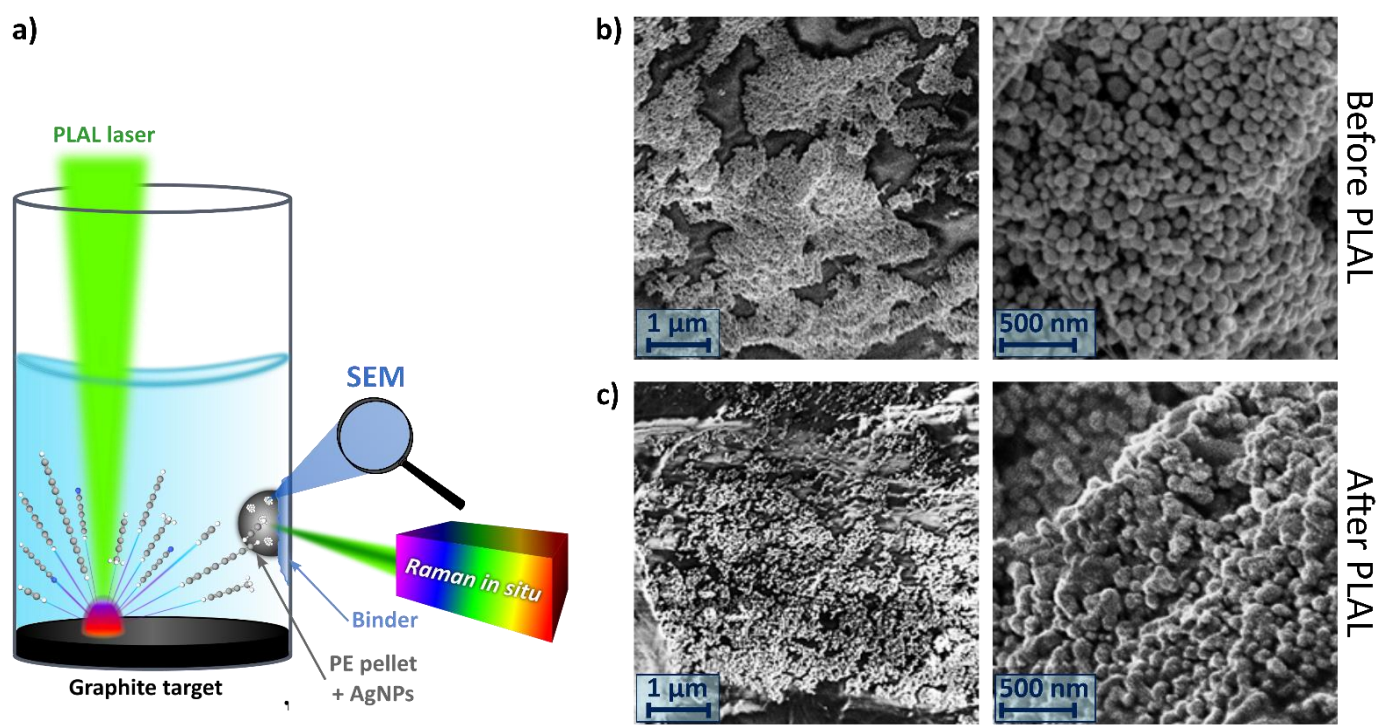

**Figure 1.** a) Scheme of *in situ* SERS setup: PE pellet functionalized with silver nanoparticles attached to the glass vial with a polyvinyl alcohol layer as the binder. The Raman laser beam is focused on the pellet surface

from outside the glass vial. The ablation laser (PLAL laser) is directed on the graphite target far away from the pellet. SEM images of the pellet before (b) and after (c) the ablation at two different magnifications.

Figure 2 shows SERS spectra continuously recorded during 60 minutes of ablation. Each spectrum represents the signal integrated over 10 acquisitions of 1 s each. The investigated spectral range covers both the sp and $sp^2$ Raman bands and some Raman peaks of the solvent (acetonitrile, ACN), namely the CC stretching mode at 920 cm$^{-1}$ and C≡N stretching mode at 2254 cm$^{-1}$ [66]. The CC stretching mode of ACN is used as an internal reference since it is not influenced by the presence of polyynes or by other carbon-based byproducts. Conversely, the intensity of the C≡N stretching mode may be influenced by monocyano-capped polyynes, produced when ablating in ACN [10,15,28]. Selected SERS spectra at specific time intervals (0, 5, 30, and 60 min) are reported in panel b) of Fig. 2. From the analysis of the spectral features, we can divide the frequency range in the $sp^2$ carbon (1000 - 1700 cm$^{-1}$), and in the sp-carbon (1800 - 2200 cm$^{-1}$) regions [20,25,43,47,52,67,68]. The $sp^2$ region is characterized by the convolution of the D + G bands characteristic of $sp^2$ carbon [20,43,68,69]. The sp-carbon region shows a broad band due to the convolution of the SERS signals of the mixture of polyynes attached to the pellet [33,43,45]. The interaction of polyynes with AgNPs on the PE pellet broadens the characteristic CC collective mode of sp chains and introduces new bands, as already observed in SERS spectra of size- and capping-selected polyynes [15,33,45,70]. Such behavior indicates a strong chemical interaction between polyynes and metal nanoparticles (i.e. chemical SERS effect) with the possible occurrence of a charge transfer [31,48,71].

The time evolution of the sp and $sp^2$ Raman features during the ablation was investigated by integrating the Raman intensity in their respective spectral range. Fig. 2c displays the time evolution of the integrated area of $sp^2$- and sp-carbon signals ($A_{sp^2}$ and $A_{sp}$, respectively). During the first minutes of ablation, both the $sp^2$ and sp areas increase at different rates and reach different maximum values. The $sp^2$ area, after touching its maximum, shows an approximately constant behavior, characterized by a feeble decrease at longer ablation times, as can be also appreciated from the signal at 60 min in Fig. 2b. The sp-carbon area, instead, reaches its maximum value at approximately 130 s and then starts a fast decay that asymptotically tends to a lower value compared to that of the $sp^2$ signal. We investigated the concentrations of H-capped polyynes by *ex situ* HPLC analyses at discrete ablation times to validate *in situ* SERS data. Indeed, we know from previous works that the majority of the sp-carbon compounds in the mixture are H-capped polyynes [15,23,70,72]. As reported in Fig. 2d, the concentration of size-selected H-capped polyynes in the solution continuously grows during the ablation process.

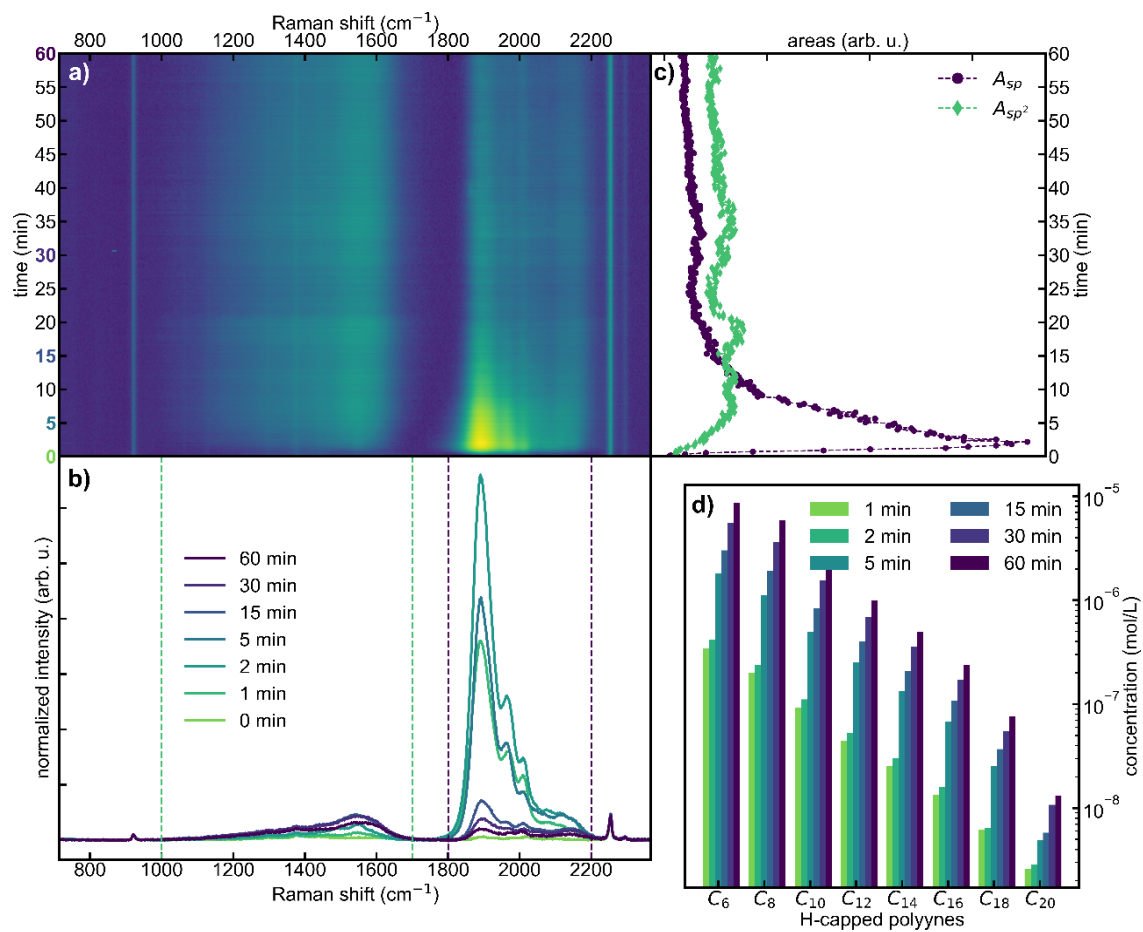

**Figure 2.** a) 2D plot of the evolution of the SERS signal during the 60 min ablation. b) SERS spectra of the species attached to the PE pellet at fixed time intervals. Dashed colored vertical lines bound the sp$^2$ (1000-1700 cm$^{-1}$) and sp (1800-2200 cm$^{-1}$) spectral regions. c) Integrated areas of the sp$^2$ ($A_{sp^2}$, 1000-1700 cm$^{-1}$) and sp ($A_{sp}$, 1800-2200 cm$^{-1}$) regions as a function of time. d) Concentration of size-selected H-capped polyynes (HC$_{2n}$H where the number of triple bonds n = 3-10, shortened in the figure as C$_{2n}$) after specific time intervals (i.e. 1, 2, 5, 15, 30, and 60 minutes), extracted from their corresponding chromatographic areas.

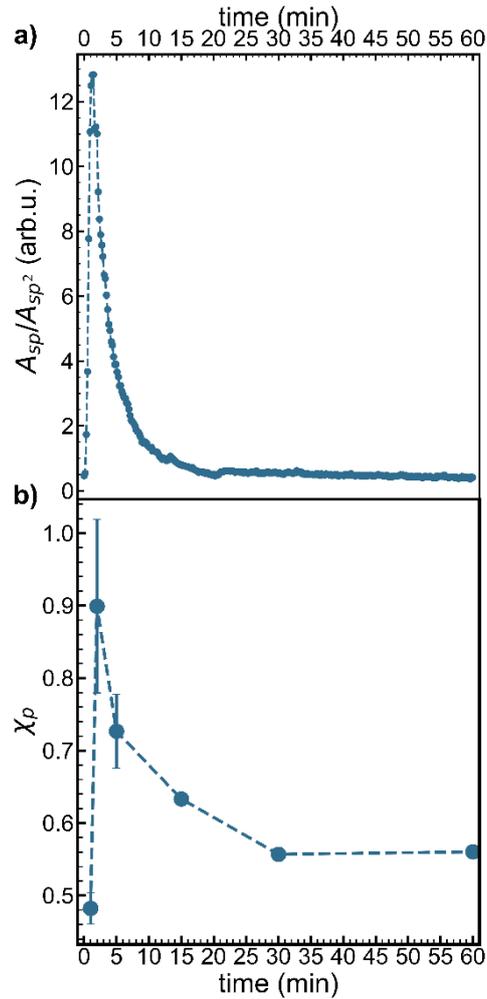

**Figure 3.** a) Ratio between the areas of sp and sp² regions as a function of time of ablation. b) Evolution of the index of purity ($\chi_p$) as a function of time, extracted from the UV-Vis spectra of Fig. S4, with the method defined in Peggiani et al. [15]

The discrepancy between the evolution in time of polyynes SERS signals and the corresponding concentration in solution suggested investigating more deeply the ablation process. We analyzed both *ex situ* UV-Vis data related to the mixture of polyynes at crucial ablation times and SERS spectra obtained by the SERS probe. We first consider the ablation efficiency in terms of polyynes production, as representative of the SERS sp signal, compared to sp² carbon species, considered from now on as byproducts. Figure 3 shows the ablation efficiency from two perspectives: in panel a), as the ratio between the integrated areas of the sp- and sp²-carbon SERS bands obtained from *in situ* SERS ($A_{sp}/A_{sp^2}$), in panel b), by calculating an index of purity of the mixture, $\chi_p$, obtained from UV-Vis absorption spectra of the solution at increasing ablation times (see Fig. S4). The index of purity is calculated as the ratio between the area of polyyne absorption peaks and the background attributed to byproducts [14,15,24,32]. In both cases, the evolution is non-monotonic. Starting from SERS data, the $A_{sp}/A_{sp^2}$ ratio rapidly increases until 109 s, then quickly falls and finally slowly decreases. The descending phase cannot imply a decline in the formation of polyynes in the solution because, from HPLC analysis (see Fig. 2d), the concentration of each H-capped polyyne grows during the whole ablation process. However, the sp SERS signal decreases as can be seen in Fig. 2c, and this can be a sign of the degradation of polyynes on AgNPs. Conversely, we did not record a proportionate increase of the sp$^2$ SERS signal that only slightly increases during the first 5 minutes after which it remains quite unchanged for the rest of the ablation (see Fig. 2c). Thus, the evolution of the $A_{sp}/A_{sp^2}$ ratio must imply the existence of a mechanism of degradation of polyynes on the pellet that contrasts the stabilization governed by the interaction with AgNPs. The evolution in time of the $A_{sp}/A_{sp^2}$ ratio led us to believe that this degradation process starts in the first stages of the ablation, at least locally on the pellet.

In this framework, we modeled the time evolution of the $A_{sp}/A_{sp^2}$ ratio of Fig. 3. The onset of a degradation mechanism observed during the first steps of ablation can be approximated with a sigmoid function $\frac{a}{1+e^{-(t+t_0)/\tau_1}}$, where $\tau_1$ represents the characteristic time of this process and $t_0$ the midpoint of the sigmoid [73–75]. The fast decrease can be rationalized with a decaying exponential as $e^{-(t+t_0')/\tau_2} + c$, where the decay time is equal to $\tau_2$, $c$ is the asymptotic value and $t_0'$ is a rigid time-shift introduced in the fitting model [21,76]. Employing this method, the calculated time constant of the rising part of Fig. 3a turns out to be $\tau_1 = 9.2 \pm 0.25$ s, while the descending phase has a decay time $\tau_2^{exp} = 173 \pm 1.66$ s. The calculated $\tau_1$ is comparable or even lower than the temporal resolution of our SERS measurement (i.e. 10 s), so the degradation starts almost synchronously with the ablation process and may be associated with crosslinking reactions with other polyynes or sp$^2$ species, even if polyynes are stabilized by AgNPs on the pellet. Indeed, crosslinking is one of the main channels of degradation of polyynes, especially in presence of a high density of carbon species, i.e. sp$^2$ byproducts or other reactive polyynes – i.e. without bulky endgroups – as may be the case of the surface of the pellet[15,52,76]. If the degradation products remain on the PE pellet, we should observe the rise of the sp$^2$ signal and the decrease of the sp one, while if they detach, both signals should lower. From the observations about the $A_{sp^2}$ area (see Fig. 2c), we can conclude that both the mechanisms occur at the same time, producing a sort of dynamical equilibrium. Similarly, in the solution, a single polyyne is surrounded by other polyynes and byproducts expelled from the ablation site with high speed that slow down and start moving in Brownian-like motion. Such systems may chemically interact with the polyynes causing their degradation. In this sense, the situation, probed by SERS, of a polyyne fixed on the PE pellet gives back the kinetics of degradation of polyynes in solution.

We analyzed the index of purity ($\chi_p$) in the solution, extracted from *ex situ* UV-Vis measurements at different ablation times (see Fig. S4), to validate the conclusions coming from our *in situ* SERS approach. The index of purity provides conceptually the same information of the SERS $A_{sp}/A_{sp^2}$ ratio, but from the perspective of the entire solution and without the interaction with silver nanoparticles. Indeed, during the first minutes of ablation, $\chi_p$ increases following the same time evolution of the *in situ* SERS probe which therefore provides the picture of what is happening in the solution.

The decrease of the SERS $A_{sp}/A_{sp^2}$ ratio and the index of purity $\chi_p$ is approximated with exponential decay in the fit model. Regarding the SERS ratio, we calculated a characteristic time $\tau_2^{exp} = 173 \pm 1.66$ s. Similarly, we estimated the decay time $\tau_2^{\chi_p}$ of the descending part of the index of purity from Fig. 3b. The calculated value, i.e. $\tau_2^{\chi_p} = 527 \pm 243$ s, despite the large fit error deriving from the small dataset available (6 points for 6 fit variables), is comparable with $\tau_2^{exp}$ extracted from SERS data. These fast decays can be rationalized considering that polyynes undergo degradation not only on the PE pellet but also in the solution.

It is reasonable to assume that the fast degradation rate in the case of *in situ* SERS can be related to the continuous generation of polyynes and byproducts that speed up the degradation of polyynes. Indeed, if we stop the ablation after 109 s (see Fig. S5), i.e. at the maximum of the SERS $A_{sp}/A_{sp^2}$ ratio extracted from Fig. 3a, we should expect a slower degradation of polyynes because the synthesis is stopped, but polyynes can still crosslink throughout the solution and on the pellet too. Applying the same fit model, the decay time when the ablation is stopped after 109 s ($\tau_2^{109\,s} = 3387 \pm 554.66$ s) is larger than in the case of continuous ablation ($\tau_2^{exp} = 173 \pm 1.66$ s), while the time constant $\tau_1$ of the activation of the degradation mechanism of polyynes remains rather unchanged ($6 \pm 1.05$ s) (see Fig. S6). We compared our data with the work of Lucotti et al.[43] in which they recorded the evolution in time of the SERS signal by adding a colloidal solution of AgNPs to a mixture of polyynes previously produced by arc discharge in methanol. We extracted the exponential decay time ($\tau_2^{lit} = 382 \pm 41.6$ s) by calculating the SERS $A_{sp}/A_{sp^2}$ ratio from Fig. 4b of ref. [43]. In that case, the degradation of polyynes was due to the increased probability of crosslinking reactions induced by the aggregating effect of the Ag colloids. Despite the differences between the two setups, the similarity of the decay times confirms crosslinking as the main channel of degradation of polyynes, even if they are stabilized by AgNPs. The reason that explains the larger decay time of Lucotti et al. ($\tau_2^{lit}$) compared to the case of continuous ablation ($\tau_2^{exp}$) is the same as discussed in the case of 109 s of ablation ($\tau_2^{109\,s}$), i.e. there is no further generation of reactive polyyne.

We further investigated this phenomenon with a dedicated experiment in which we transferred a PE pellet, covered by polyynes and byproducts produced during 60 min of ablation, in pure acetonitrile solution, and we track its status via *in situ* SERS. In this case, $A_{sp}$ and $A_{sp^2}$ decrease with a comparable slope (see Fig. S7). The decrease of $A_{sp^2}$ is related to the desorption of byproducts from the PE pellet, confirmed by *ex situ* UV-Vis spectrum in which we detected a signal similar to the absorption background of byproducts in Fig. S1 and S4. We concluded that the detachment of byproducts is favored by improved mobility gained from the local heating of AgNPs that absorb the Raman laser during SERS measurements. The decrease of $A_{sp}$ cannot be connected to crosslinking since there are not enough reactive carbon species in the solution. Indeed, as already demonstrated, polyynes do not desorb from the PE pellet and thus the decrease of $A_{sp}$ is caused by a local thermal degradation of polyynes on the pellet due to the Raman-induced heating of AgNPs. The comparison of the SERS $A_{sp}/A_{sp^2}$ ratio in the three cases (see Fig. S6), 60 min or 109 s of ablation time and without ablation, showed that the rate of degradation increases as the probability of crosslinking reactions increases, once we assumed that the local thermal degradation induced by the Raman laser does not change since the measurement parameters remain fixed in all the three experiments.

We here showed that our approach can provide *in situ* time-resolved information about the processes occurring during the synthesis of polyynes by PLAL without affecting the chemical and physical environment of the formation of polyynes (e.g. by adding colloidal nanoparticles). Indeed, we experimentally demonstrated that the formation and the degradation (i.e. crosslinking) of polyynes occur simultaneously during PLAL and not only after their synthesis, as already reported in the literature [4,15,21,43,76]. Moreover, this method allows the evaluation of the time evolution of the efficiency of polyynes production compared to byproducts, without the delay or the perturbation of *ex situ* analyses.

## 3.2 Formation and degradation mechanisms of long and short polyynes

Here we focus on monitoring the role of the length of polyynes on their formation and dynamics during the ablation. The SERS spectrum already in the first minutes of ablation (Fig. 2b) exhibits a clear asymmetrical distribution of features in the polyyne region with a larger intensity at the low-frequency edge of the sp-carbon spectral region (1800-2200 cm$^{-1}$). This suggests that long polyynes may have a primary role in the response of the *in situ* SERS probe. Indeed, it is well known that the ECC mode of H-capped polyynes redshifts and its Raman activity grows (see Fig. S8 in SI) as the chain length increases [1,31,33,45,77]. Even though the interaction of polyynes with silver nanoparticles introduces low-frequency bands and prevents an easy interpretation of the SERS spectra [33,43,45,70], we analyzed the spectra to extract information on the behavior of polyynes of different lengths. We developed a custom algorithm to deconvolve the overall polyyne band at each time instant of the *in situ* SERS measurement. The fit model was built as a linear combination of experimental SERS spectra of size-selected H-capped polyynes, reported in Fig. 4a, corrected by a weighting factor that represents the fraction of every single chain at a specific time. After applying the fitting algorithm, the weighted SERS signal of each H-capped polyyne has been divided by its corresponding ECC Raman mode activity computed with DFT calculations (see Fig. S8 in the SI). In such a way, we corrected our estimations from the nonlinear growth of the ECC Raman mode activity of polyynes with increasing chain length. Since the ECC intensity always prevails over the other collective modes of polyynes in the spectral region selected (1800-2200 cm$^{-1}$), as a first approximation, we have assumed its Raman activity alone as the corrective factor [31,77].

The SERS spectra of size-selected H-terminated polyynes with increasing length, from 6 to 20 carbon atoms were collected employing PE pellets sunk in an acetonitrile/water mixture of size-selected polyynes collected by HPLC. We grouped polyynes in long and short chains, depending on the intensity of their SERS signal below or above 2000 cm$^{-1}$. Indeed, the hallmarks in the spectrum of long polyynes are the merging of the high- and low-frequency bands and a negligible signal above 2000 cm$^{-1}$ [33]. Thus, as can be deduced from Fig. 4, short polyynes include chains from HC$_6$H till HC$_{12}$H, and long polyynes range from HC$_{14}$H to HC$_{20}$H. Figure 4b shows some results of the fitting procedure at specific times of ablation. The contribution of the SERS signal of each H-capped polyyne is displayed as well to highlight the different magnitudes both in frequency and time. The model suffers from some overfitting since all the SERS spectra employed in the fit possess one or more bands below 2000 cm$^{-1}$. Moreover, the fit slightly overestimates the intensity of the peak at 1950 cm$^{-}$

[1] at 5 min ablation time, while it does not fit well the tail at higher frequencies at longer times. These mismatches may be due to the choice to consider in the model H-capped polyynes starting from 6 atoms of carbon because they are the most concentrated species in the mixture. However, it is probable that shorter chains, namely $HC_2H$ and $HC_4H$, have been produced during the ablation and may bring significant contributions to the SERS band from 2100 to 2200 cm$^{-1}$. However, we did not consider these compounds in our analysis because of the lack of UV-Vis spectra in literature and the impossibility to detect and separate them with our HPLC apparatus (their absorption maxima are below 190 nm that is the lower limit of the DAD detector of HPLC). Moreover, we examined only H-capped polyynes, but other terminations exist, as cyano and methyl endgroups (see Peggiani et al. [15]), as well as unknown-capped sp-carbon chains that may have a non-negligible SERS signal. Nevertheless, the selected H-capped polyynes represent the most significant set of sp-carbon chains in the mixture and allow us to evaluate the formation of polyynes based on their length without introducing too many overfitting errors. Further developments are needed to include all the polyynic contributions and to correctly evaluate the evolution of the single chain.

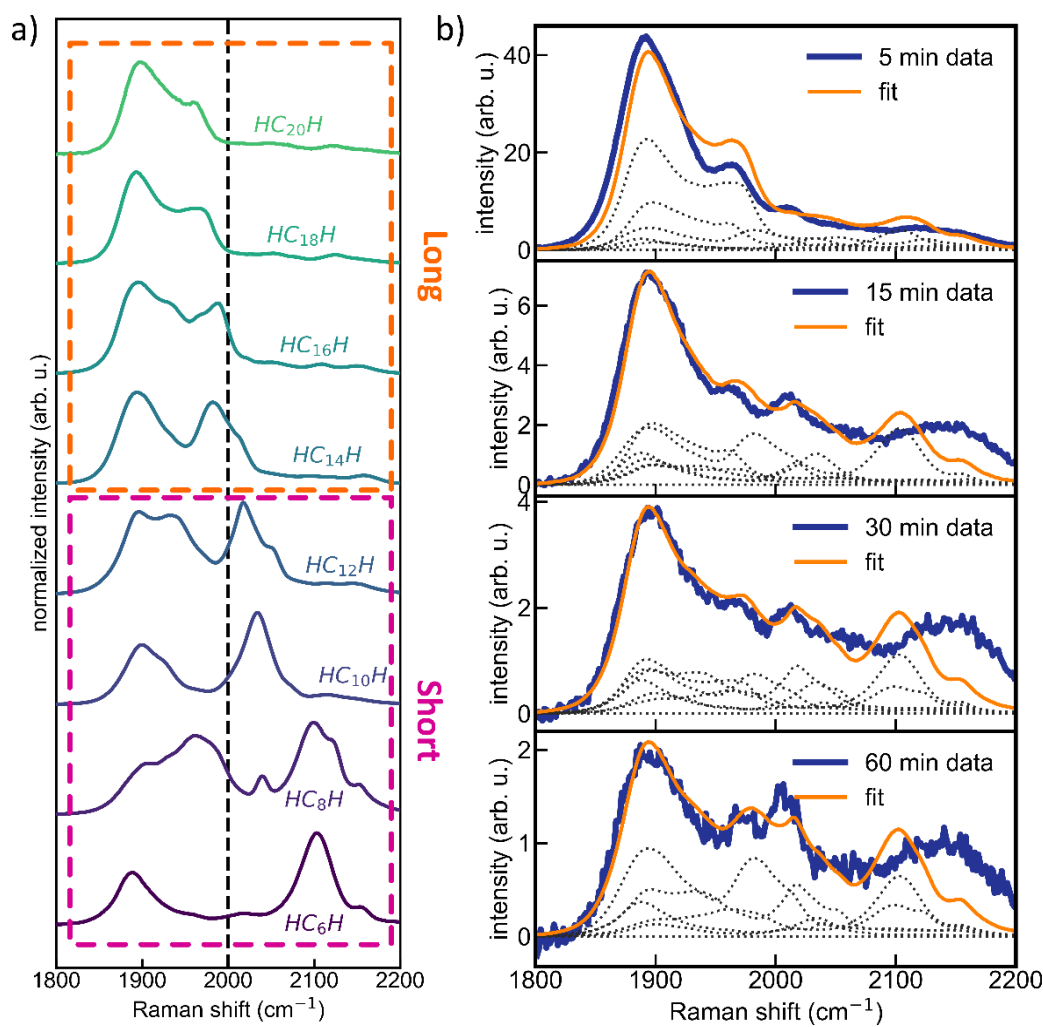

**Figure 4.** a) SERS spectra of size-selected H-capped polyynes probed employing a pristine PE pellet with AgNPs. Each spectrum is normalized to its maximum between 1800 and 2200 cm$^{-1}$. The dashed line at 2000 cm$^{-1}$ helps to differentiate between short and long chains. b) Deconvolution of *in situ* SERS signals of sp region at different times (maximum ratio long/short at 109 s, 5, 15, 30, 60 mins) by the model fit. The experimental data (solid blue line) is compared to the calculated fit function (solid orange line). The different contributes of single polyynes are reported for each time with a dotted gray line.

Employing our SERS probe, we analyze the evolution of the synthesis of long and short polyynes to catch possible differences depending on the chain length. The SERS $A_{long}/A_{short}$ ratio, reported in Figure 5a, represents the ratio between the SERS areas of long and short polyynes. The influence of the length of polyynes anchored to the pellet on their kinetics is compared with the conceptually identical dynamics in the solution (Fig. 5b), extracted from *ex situ* HPLC analysis at specific ablation times. For the reasons discussed before, we choose to group polyynes in long and short chains, depending on their activity below or above 2000 cm$^{-1}$.

Considering the SERS $A_{long}/A_{short}$ ratio, its maximum sets at 32 s, while the characteristic times are $\tau_1 = 5.75 \pm 0.45$ s and $\tau_2 = 185 \pm 4.3$ s. The maximum of $A_{long}/A_{short}$ ratio, extracted from HPLC data, is found at 120 s, and the characteristic times are $\tau_1 = 4 \pm 3.5$ s and $\tau_2 = 1562 \pm 15.8$ s. Even if, the scarcity of data points of the *ex situ* analysis makes it complicated to have a precise estimation of the time constants, overall, the two trends have similar shapes and quite comparable scales.

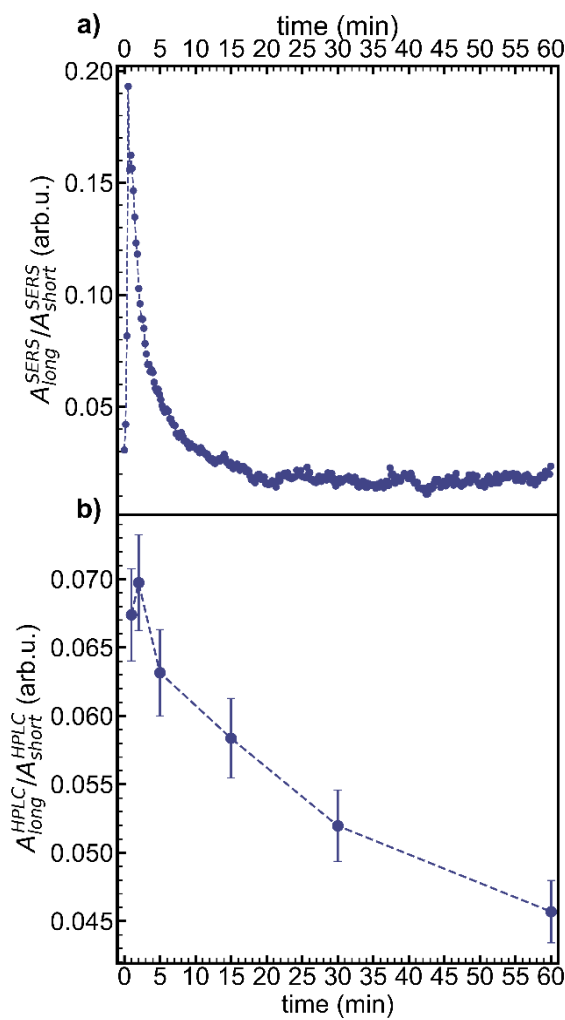

**Figure 5.** a) Evolution in time of the ratio of SERS areas of long *versus* short polyynes. b) Evolution in time of the ratio of the chromatographic areas of long over short polyynes, detected by HPLC measurements at specific ablation times (1, 2, 5, 15, 30, and 60 minutes).

The evolution in time of the $A_{long}/A_{short}$ ratio extracted from *in situ* SERS suggests that the growth of long polyynes is favored at short times while at longer times short polyynes are predominant. To interpret these data, some points need to be considered. As already known for π conjugated materials, the reactivity of polyynes scales up with their length, therefore a different degradation rate should be considered [1,31,78]. However, the PLAL technique has already been demonstrated to be a proficient ambient for the formation of long polyynes up to $HC_{30}H$ in non-polar solvents [26]. Based on these considerations, we can assume that at least in the first instants of ablation the synthesis rate of longer polyynes exceeds that of shorter ones. This is indeed supported by our results, showing that the $A_{long}/A_{short}$ ratio of chromatographic areas reported in Fig. 5b increases from 1 to 2 minutes of ablation. The subsequent decrease of the $A_{long}/A_{short}$ ratio both in solution (i.e. from *ex situ* HPLC) and on the PE pellet (i.e. from *in situ* SERS) could be explained only if long polyynes degrade faster than shorter ones. Indeed, we exclude that the synthesis process of long and short polyynes changes during the ablation. Such variation could be ascribed to laser-induced photodegradation processes or a variation of the liquid environment around the plasma plume (e.g. thermal degradation). Considering the laser-induced photodegradation processes, polyynes and $sp^2$ carbon byproducts dissolved in the solution can have different interactions with the ablation laser. Nevertheless, even the longest polyyne produced by PLAL in acetonitrile cannot absorb the ablation laser (maximum absorption wavelength for $HC_{22}H$ is at 362 nm) and undergo photo-induced degradation. Byproducts generated during PLAL, instead, feature a non-negligible absorption in the visible light, as it has been demonstrated by the long tail in the UV-Vis absorption spectra of the mixture of polyynes obtained by PLAL (see also Fig. S1 and S4 in the SI) [14,15,32]. Those byproducts suspended in the solution can be fragmented into $C_2$ radicals and contribute to the growth of polyynes, similarly to the role of graphite particles or C60 dispersed in solution discussed in the works of Tsuji et al. [23,72]. Nevertheless, the concentration of byproducts produced by PLAL is much lower than the amount of graphite particles ($4.2 \cdot 10^{-2}$ mol/L [72]) or C60 ($3.8 \cdot 10^{-3}$ mol/L [23]) employed in previous experiments as the main source of carbon atoms for polyyne synthesis. This value is slightly larger than the concentration of polyynes from the calculation of $\chi_p$ in Fig. 3b or by looking at the Raman spectrum in the solvent of Fig. S2a where the $sp^2$-related Raman signal is negligible. So, we consider that the contribution of byproducts does not boost the synthesis of polyynes in a relevant way. Regarding the chemical environment around the plasma plume, it may be perturbed by partial substitution of solvent molecules by sp-carbon chains or byproducts that can act as secondary carbon sources and may induce alterations in the size distribution of synthesized polyynes. Indeed, if carbon species remain in the proximity of the laser plume, they can degrade due to the high-temperature gradients experienced at the plasma-liquid boundary. However, if we assume that the majority of polyynes and byproducts are pushed away from the ablation site due to shockwaves generated by the collapse of the cavitation bubble produced by the quenching of the plasma plume, we expect that the thermal degradation process can be neglected [58,62,63]. Once we have excluded the photoinduced and thermal degradations, crosslinking reactions candidate as the main channel of degradation of polyynes during the ablation and justify the faster decrease of longer polyynes because of their higher reactivity. Indeed, the SERS $A_{long}/A_{short}$ ratio in the case of 109 s of ablation features a much slower decrease than that of continuous ablation (see Fig. S9) due to a lower concentration of diluted carbon species in the solution that depresses crosslinking reactions and raises the lifetime of long polyynes both on the PE pellet and within the solution.

Thus, *in situ* SERS data allows us also to keep track of the evolution of the distribution of polyyne length during the ablation and understand the reasons behind the final distribution of lengths, which is biased towards shorter chains. From our observations, we can exclude that the different degradation rates between short and long polyynes are imputed to photoinduced degradation caused by the ablation laser or thermal degradation provoked by the extreme conditions in the plasma plume. Conversely, the crosslinking probability increases according to the chain length, i.e. longer polyynes possess more radical-like sites, a well-known property of π-conjugated molecules. Therefore, longer chains lose their linearity and reorganize into $sp^2$ species faster than shorter wires, inducing the observed size distribution of a mixture of polyynes synthesized by laser ablation.

## Conclusions

We employed a novel *in situ* SERS approach to investigate the formation of polyynes by pulsed laser ablation in liquid acetonitrile. The sensitivity given by the SERS effect granted us to study polyynes formation with an unprecedented time-resolution of only a few seconds. It is also possible to acquire information even when the concentration of polyynes is too small for conventional Raman (≤$10^{-8}$ mol/L), i.e. in the early stages of ablation and for longer chains. Another advantage of the SERS approach based on a solid PE pellet functionalized with AgNPs is that the chemical and physical environment of the ablation is not affected by SERS-active nanostructures dispersed in the solvent. Thus, this SERS probe can provide reliable information to *in situ* and *in operando* observe the formation process of sp-carbon chains avoiding the delay and perturbation of polyynes due to *ex situ* characterization measurements.

This methodology allowed us to analyze the dynamics of the synthesis of polyynes: we show that the degradation processes (crosslinking reactions) start almost synchronously with the formation of sp-carbon chains and we verified that they depend on the concentration of both polyynes and byproducts in the mixture. Moreover, we have analyzed by *in situ* SERS how the dynamics of short and long sp-carbon chains evolve during the ablation time. By looking at the pellet, we observed a majority of shorter chains during the ablation process. Indeed, thanks to their smaller surface area, carbon species adsorbed on the pellet can interact less often with sp-carbon chains and byproducts dissolved in the solution. Thus, the degradation affects preferentially longer polyynes and the probability of crosslinking grows as the chain length increases. Thus, the final distribution in the solution at the end of the ablation is unbalanced towards shorter chains. *Ex situ*

UV-Vis and HPLC measurements confirmed the information gained through *in situ* SERS data, confirming that the dynamics on the pellet simulate what is happening in the solution.

Even though we only investigated synthesis in acetonitrile, *in situ* SERS methodology can be applied to any liquid or gaseous environment. We assume that the conclusions we have obtained about the synthesis of polyynes maintain their generality also with other solvents. However, the thermodynamical (e.g. viscosity and diffusivity) and the physicochemical (e.g. C/H ratio, radical production, and absorption of the laser beam) properties of the solvent may influence the efficiency of the synthesis. Moreover, we expect that the dynamics of crosslinking reactions, represented by the characteristic time constants, should change and indicate the most efficient environment for the synthesis of polyynes. Our SERS approach can contribute to shed light on the formation mechanisms of sp-carbon chains and of carbon nanostructures during laser ablation in liquid and give information to improve the synthesis efficiency toward the exploitation of carbon atomic wires in technological applications.

## Acknowledgements
The authors thank A. Vidale for his assistance during the development and the measurements with in situ SERS, and fruitful discussions about the data. The authors acknowledge funding from the European Research Council (ERC) under the European Union's Horizon 2020 research and innovation program ERC-Consolidator Grant (ERC CoG 2016 EspLORE grant agreement no. 724610, website: www.esplore.polimi.it).

# Supporting Information (SI)

## Transfer matrix method applied to the calculation of laser fluence

The transfer-matrix method was applied to calculate the actual laser spot on the graphite target and thus evaluate the laser fluence. This method allows us to describe each element down to the optical path of the PLAL laser beam as a 2x2 matrix. The system matrix $M_s$ is then the dot product of all the matrices composing the optical path.

$$M_s = M_N \cdot \ldots \cdot M_2 \cdot M_1 \quad \text{(Eq. 1)}$$

The optical path employed in our experiments is described by the focusing lens ($M_{lens}$), the space between the lens and the liquid meniscus ($M_{air}$), the liquid meniscus ($M_{meniscus}$) and the liquid layer over the target ($M_{liquid}$). The matrices are reported below.

$$M_{lens} = \begin{bmatrix} 1 & 0 \\ -\frac{1}{f} & 1 \end{bmatrix} \quad M_{air} = \begin{bmatrix} 1 & \frac{z_{air}}{n_{air}} \\ 0 & 1 \end{bmatrix}$$

$$M_{meniscus} = \begin{bmatrix} 1 & 0 \\ \frac{n_{air} - n_{liquid}}{n_{liquid} \cdot R_c} & \frac{n_{air}}{n_{liquid}} \end{bmatrix} \quad M_{liquid} = \begin{bmatrix} 1 & \frac{h_{liquid}}{n_{liquid}} \\ 0 & 1 \end{bmatrix}$$

where $f$ is the focal length of the focusing lens, $z_{air}$ is the distance between the lens and the liquid meniscus, $n_{air}$ and $n_{liquid}$ is the refractive index of air and the liquid, respectively, $R_c$ is the radius of curvature of the liquid meniscus and $h_{liquid}$ is the thickness of the liquid layer above the graphite target. In our case, the focal length is 200 mm, while $h_{liquid}$ is 24.73 mm and $z_{air}$ corresponds to 135.27 mm. The radius of curvature $R_c$ was extrapolated from the measurement of the contact angle and resulted to be 10.79 mm. The laser beam parameters at the end of the optical path can be calculated as

$$\begin{bmatrix} r_f \\ \theta_f \end{bmatrix} = M_s \begin{bmatrix} r_0 \\ \theta_0 \end{bmatrix} \quad \text{(Eq. 2)}$$

where $r_0, \theta_0$ and $r_f, \theta_f$ are the initial and final size and angle of the laser beam, respectively. Thus, the spot size on the top of the graphite pellet results as

$$r_f = \sqrt{\frac{M_{s_{01}}^2 \lambda^2 + \pi^2 r_0^4 M_{s_{00}}}{\pi^2 r_0^4 \det(M_s)}} \quad \text{(Eq. 3)}$$

where $M_{s_{00}}, M_{s_{01}}$ are the first and second elements in the first row of the global transfer matrix $M$, respectively. Using Eq. 3 we obtain a final radius of 2.06 mm that gives a fluence of 0.37 J/cm$^2$ with an energy of 50 mJ/pulse. The size of the spot radius was also measured experimentally with a photo-sensitive paper at different conditions. We evaluate the goodness of the model calculating the R-squared parameter that resulted to be 0.933. Thus, we accept the goodness of this model for the calculation of fluence. Indeed, using geometrical optics, we obtain a different set of spot sizes that result in an R-squared of 3.04, indicating that this model completely misses the real dimensions of the experiment.

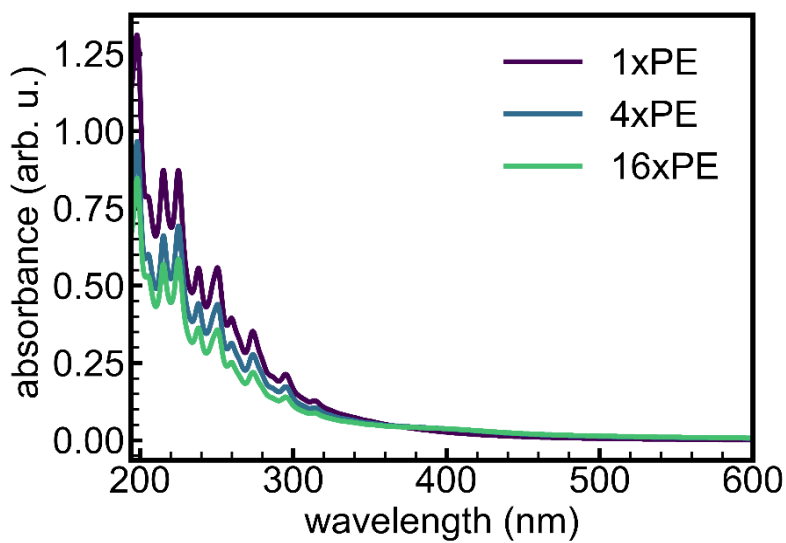

**Figure S1.** UV-Vis absorption spectra of the mixtures of polyynes ablation with an increasing number of pellets attached to the glass vial. The ablation conditions have been taken fixed to that described in Section 2.

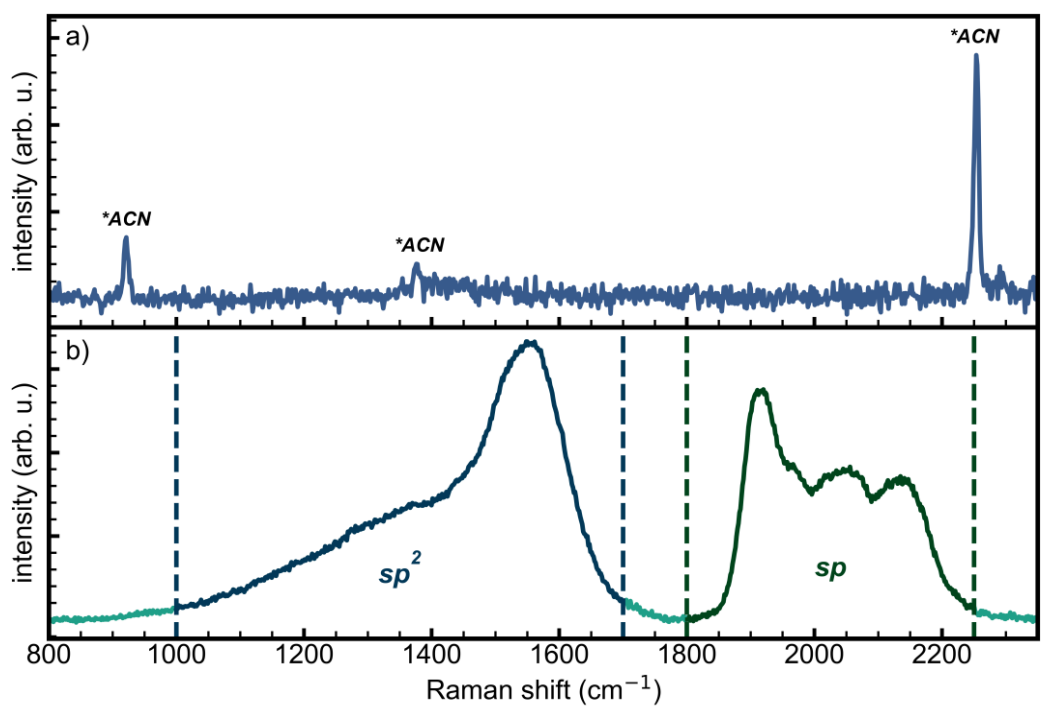

**Figure S2.** a) Raman spectrum of the mixture after 60 min of ablation by focusing the Raman laser inside the liquid contained in the ablation vial. Only vibrational modes of acetonitrile are visible. b) SERS spectrum of the PE pellet extracted from the liquid and dried in air.

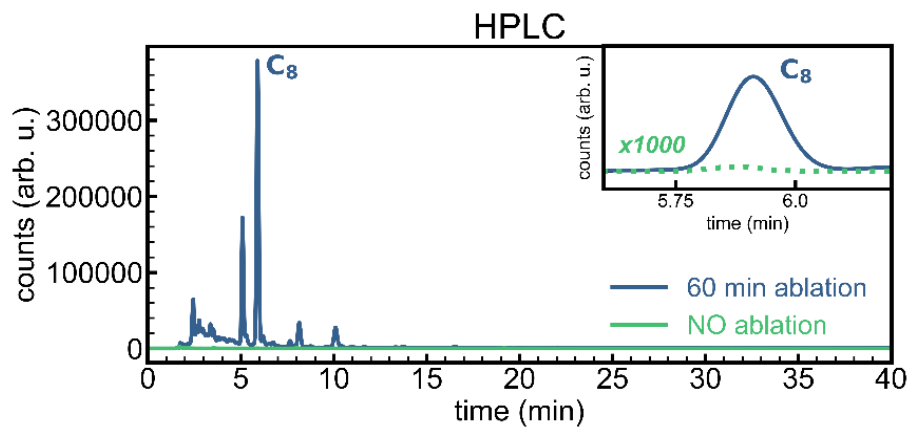

**Figure S3.** Comparison between HPLC chromatograms acquired at 226 nm (absorption maximum of $C_8$) in the cases of 60 min of ablation and without ablation (i.e. the PE pellet covered by polyynes and byproducts sunk in pure acetonitrile). The inset shows a zoom around the retention time of $C_8$.

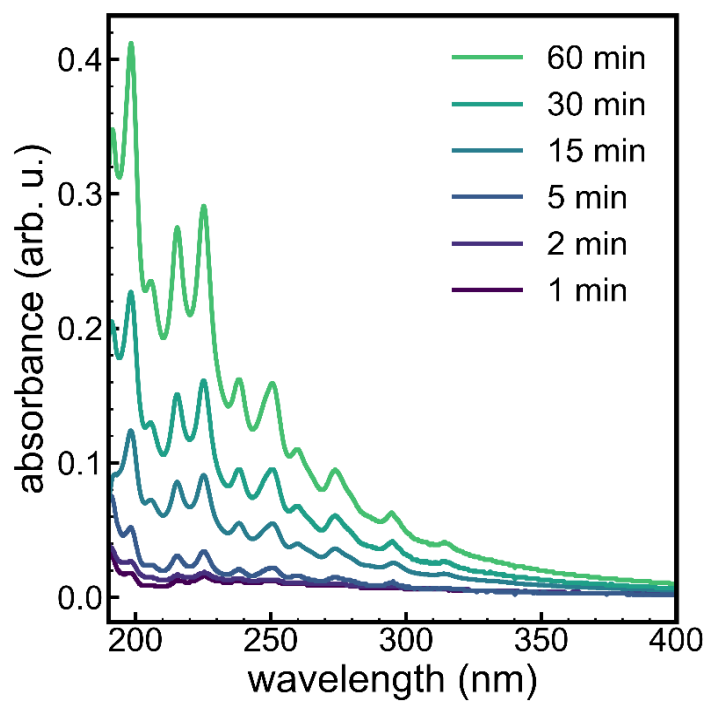

**Figure S4.** UV-Vis spectra of a mixture of polyynes after specific ablation times (1, 2, 5, 15, 30, and 60 min).

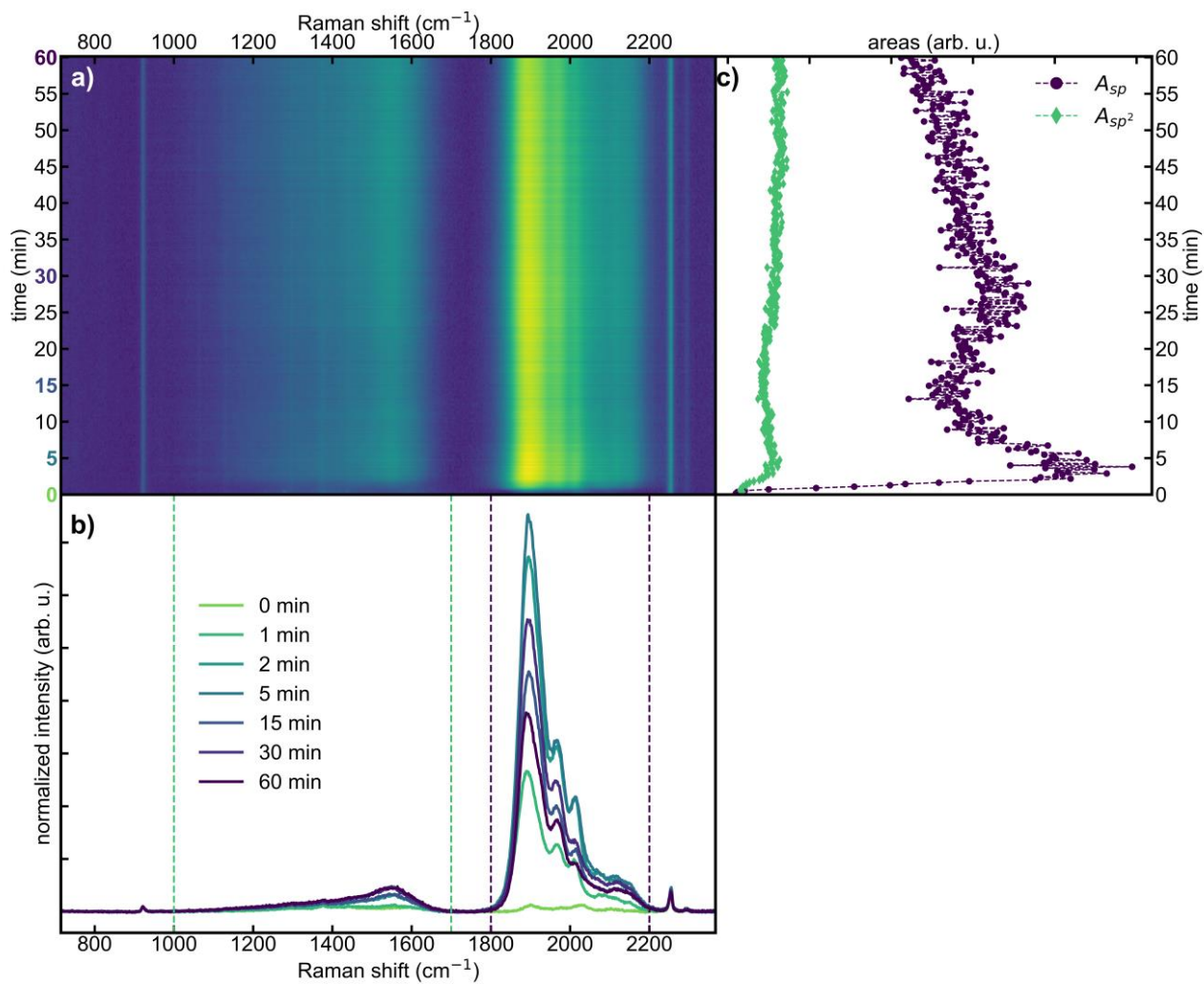

**Figure S5.** a) 2D plot of the evolution of the SERS signal during 60 min of measurement, while ablation laser was stopped after 109 s. b) SERS spectra of the species attached to the PE pellet at fixed time intervals. Dashed colored vertical lines bound the sp$^2$ (1000-1700 cm$^{-1}$) and sp (1800-2200 cm$^{-1}$) spectral regions. c) Integrated areas of the sp$^2$ ($A_{sp^2}$, 1000-1700 cm$^{-1}$) and sp ($A_{sp}$, 1800-2200 cm$^{-1}$) regions as a function of time.

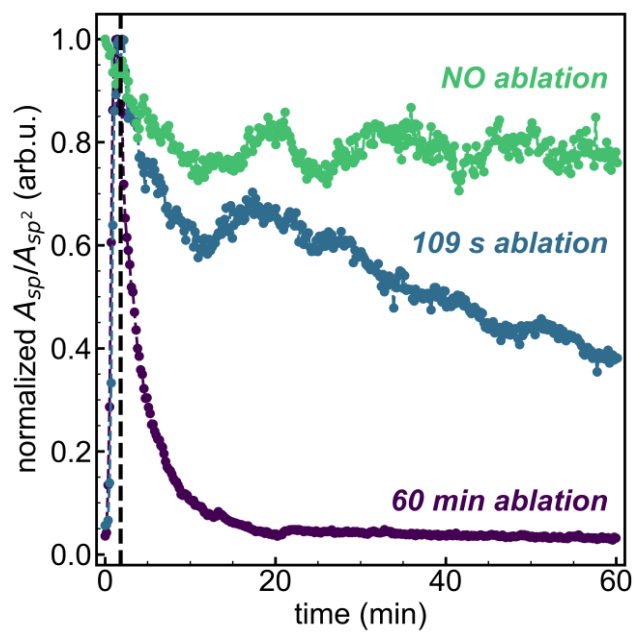

**Figure S6.** Evolution of the SERS sp/sp² ratio in the case of continuous ablation for 60 min (Fig. 2 in the main text) compared to the experiment with 109 seconds of ablation (Fig. S5) and a PE pellet sunk in pure acetonitrile (NO ablation, Fig. S7).

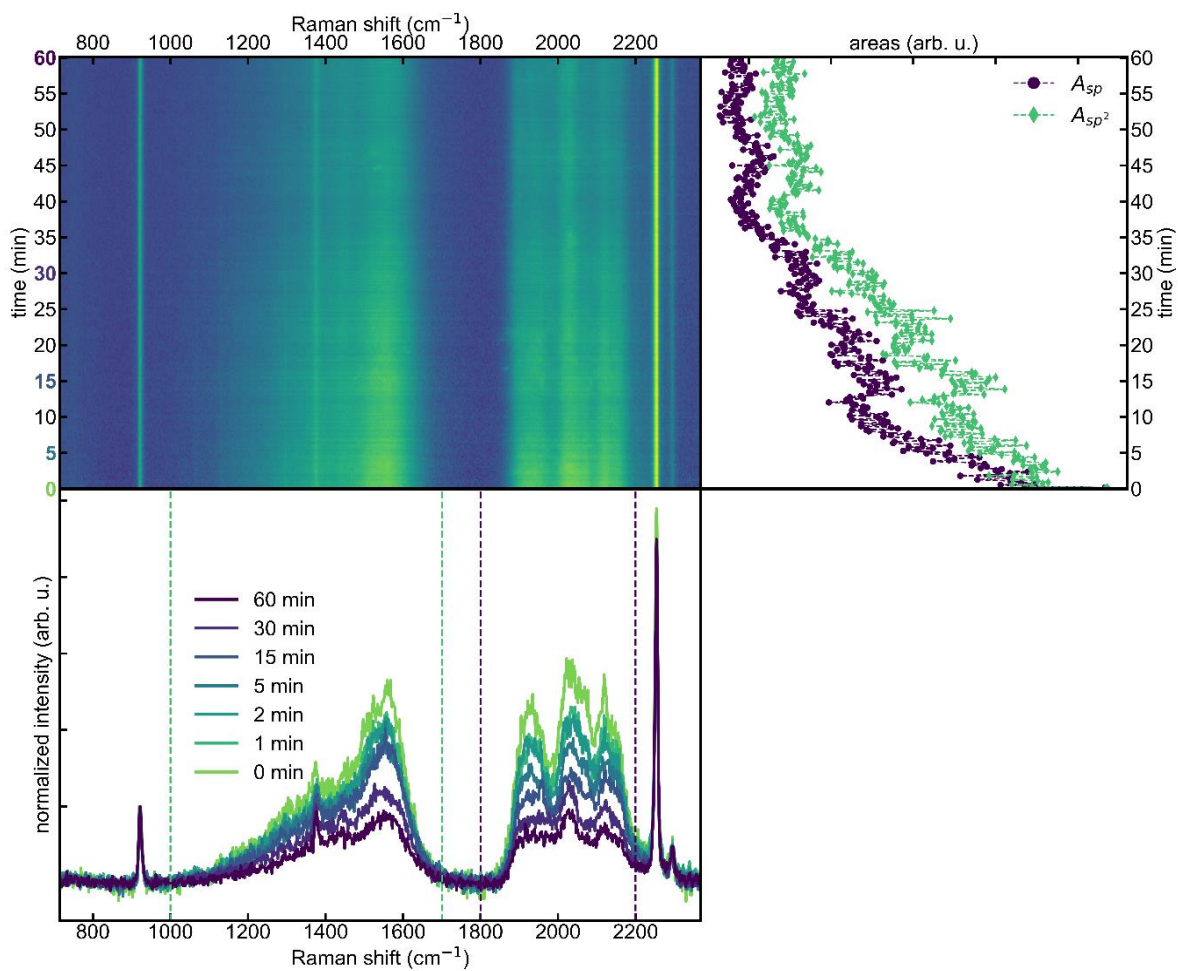

**Figure S7.** a) 2D plot of the evolution of the SERS signal during 60 min of measurement, without ablation. Polyynes and byproducts on the PE pellet derive from a previous *in situ* SERS experiment (with ablation). b) SERS spectra of the species attached to the PE pellet at fixed time intervals. Dashed colored vertical lines bound the sp² (1000-1700 cm⁻¹) and sp (1800-2200 cm⁻¹) spectral regions. c) Integrated areas of the sp² ($A_{sp^2}$, 1000-1700 cm⁻¹) and sp ($A_{sp}$, 1800-2200 cm⁻¹) regions as a function of time.

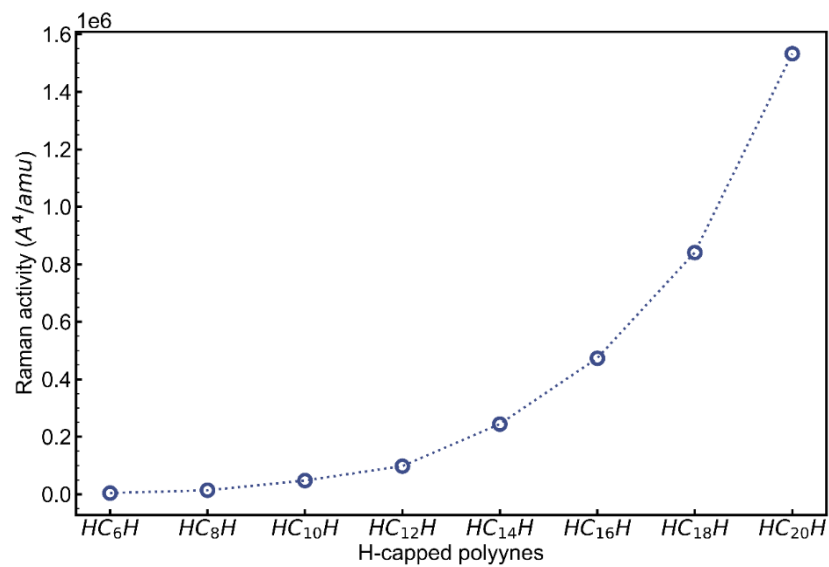

**Figure S8.** Predicted Raman activities of a series of H-capped polyynes from DFT simulations.

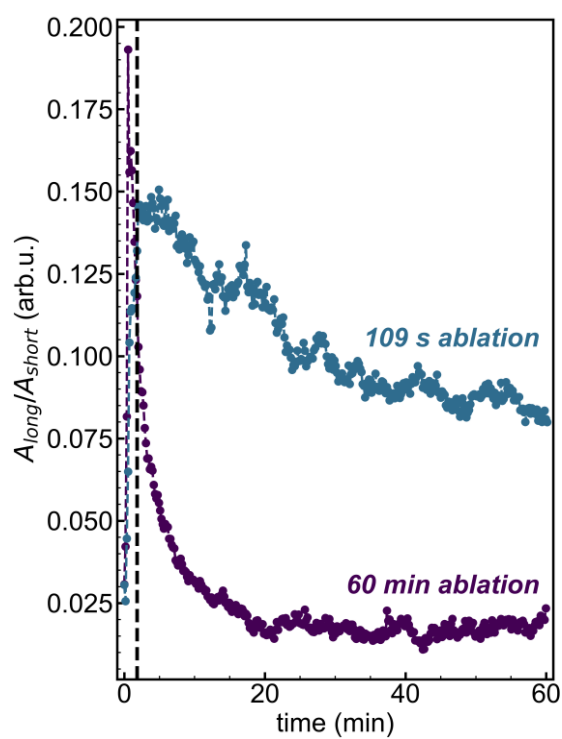

**Figure S9.** Evolution of the SERS areas of long *versus* short polyynes in the case of continuous ablation for 60 min (Fig. 2 in the main text) compared to the experiment with 109 seconds of ablation (Fig. S5).